\pdfoutput=1
\documentclass[sigconf]{acmart}

\AtBeginDocument{%
  }

\usepackage{xspace} 
\usepackage{dutchcal}
\usepackage{subfigure}
\usepackage{CJKutf8}
\usepackage{array}
\usepackage{multirow} 

\usepackage{fancyhdr}
\copyrightyear{2023}
\acmYear{2023}
\setcopyright{acmlicensed}
\acmConference[MM '23]{Proceedings of the 31st ACM International Conference on Multimedia}{October 29-November 3, 2023}{Ottawa, ON, Canada}
\acmBooktitle{Proceedings of the 31st ACM International Conference on Multimedia (MM '23), October 29-November 3, 2023, Ottawa, ON, Canada}
\acmPrice{15.00}
\acmDOI{10.1145/3581783.3613750}
\acmISBN{979-8-4007-0108-5/23/10}

\newcommand{\name}{\texttt{DiVa}\xspace}
\newcommand{\rfig}{Figure~\ref}
\newcommand{\gold}[1]{``\textbf{\textcolor[RGB]{122,98,14}{#1}}''}

\newcommand{\newsuper}[1]{``\textbf{\textcolor[RGB]{52,78,109}{#1}}''}

\newcommand{\plabel}[1]{``\textbf{\textcolor[RGB]{89,100,50}{#1}}''}
\newcommand{\badlabel}[1]{``\textbf{\textcolor[RGB]{76,55,107}{#1}}''}
\newcommand{\tnew}[1]{{\textcolor[RGB]{0,0,0}{#1}}}
\newcommand{\tgold}[1]{\textbf{\textcolor[RGB]{122,98,14}{#1}}}
\newcommand{\tfull}[1]{\textbf{\textcolor[RGB]{243,164,71}{#1}}}

\makeatletter
\def\paragraph{\@startsection{paragraph}{4}%
  {0pt}{6pt plus 2pt minus 1pt}{-1em}%
  {\normalfont\bfseries\emph}}
\makeatother
\usepackage[ruled,linesnumbered]{algorithm2e}
\usepackage{enumitem}
\setitemize[1]{itemsep=0pt,parsep=0pt,topsep=2pt,leftmargin=12pt}
\copyrightyear{2023}
\acmYear{2023}
\setcopyright{acmlicensed}
\acmConference[MM '23]{Proceedings of the 31st ACM International Conference on Multimedia}{October 29-November 3, 2023}{Ottawa, ON, Canada}
\acmBooktitle{Proceedings of the 31st ACM International Conference on Multimedia (MM '23), October 29-November 3, 2023, Ottawa, ON, Canada} \acmPrice{15.00}
\acmDOI{10.1145/3581783.3613750} \acmISBN{979-8-4007-0108-5/23/10}

\begin{document}
\title{\name : An Iterative Framework to Harvest More Diverse and Valid Labels from User Comments for Music}

\author{Hongru Liang}
\affiliation{%
   \institution{Sichuan University}
  \city{Chengdu}
  \country{China}
}
\email{lianghongru@scu.edu.cn}

\author{Jingyao Liu}
\affiliation{%
    \institution{Sichuan University}
  \city{Chengdu}
  \country{China}
}
\email{liujingyao@stu.scu.edu.cn}

\author{Yuanxin Xiang}
\affiliation{%
   \institution{Sichuan University}
  \city{Chengdu}
  \country{China}
}
\email{whutxyx@gmail.com}

\author{Jiachen Du}
\affiliation{%
  \institution{Tencent Music Entertainment}
  \city{Shenzhen}
  \country{China}
}
\email{jacobdu@tencent.com}

\author{Lanjun Zhou}
\affiliation{%
  \institution{Tencent Music Entertainment}
  \city{Shenzhen}
  \country{China}
}
\email{jedzhou@tencent.com}

\author{Shushen Pan}
\affiliation{%
  \institution{Tencent Music Entertainment}
  \city{Shenzhen}
  \country{China}
}
\email{samsonpan@tencent.com}

\author{Wenqiang Lei}\authornote{Corresponding author.}
\affiliation{%
    \institution{Sichuan University}
  \city{Chengdu}
  \country{China}
}
\email{wenqianglei@scu.edu.cn}

\renewcommand{\shortauthors}{Liang, et al.}
\begin{abstract}
Towards sufficient music searching, it is vital to form a complete set of labels for each song. However, current solutions fail to resolve it as they cannot produce diverse enough mappings to make up for the information missed by the gold labels. Based on the observation that such missing information may already be presented in user comments, we propose to study the automated music labeling in an essential but under-explored setting, where the model is required to harvest more diverse and valid labels from the users' comments given limited gold labels. To this end, we design an iterative framework (DiVa) to harvest more $\underline{\text{Di}}$verse and $\underline{\text{Va}}$lid labels from user comments for music. The framework makes a classifier able to form complete sets of labels for songs via pseudo-labels inferred from pre-trained classifiers and a novel joint score function. The experiment on a densely annotated testing set reveals the superiority of the \name over state-of-the-art solutions in producing more diverse labels missed by the gold labels. We hope our work can inspire future research on automated music labeling. 

\end{abstract}

\ccsdesc[500]{Information systems~Music retrieval}
\ccsdesc[500]{Computing methodologies~Information extraction}

\keywords{Automated Music Labeling, User Comments, Music Searching}
\maketitle
\vspace*{-1em}
\section{Introduction}
\label{sec:intro}
\begin{figure*}[t]
    \centering
    \includegraphics[width=0.8\textwidth]{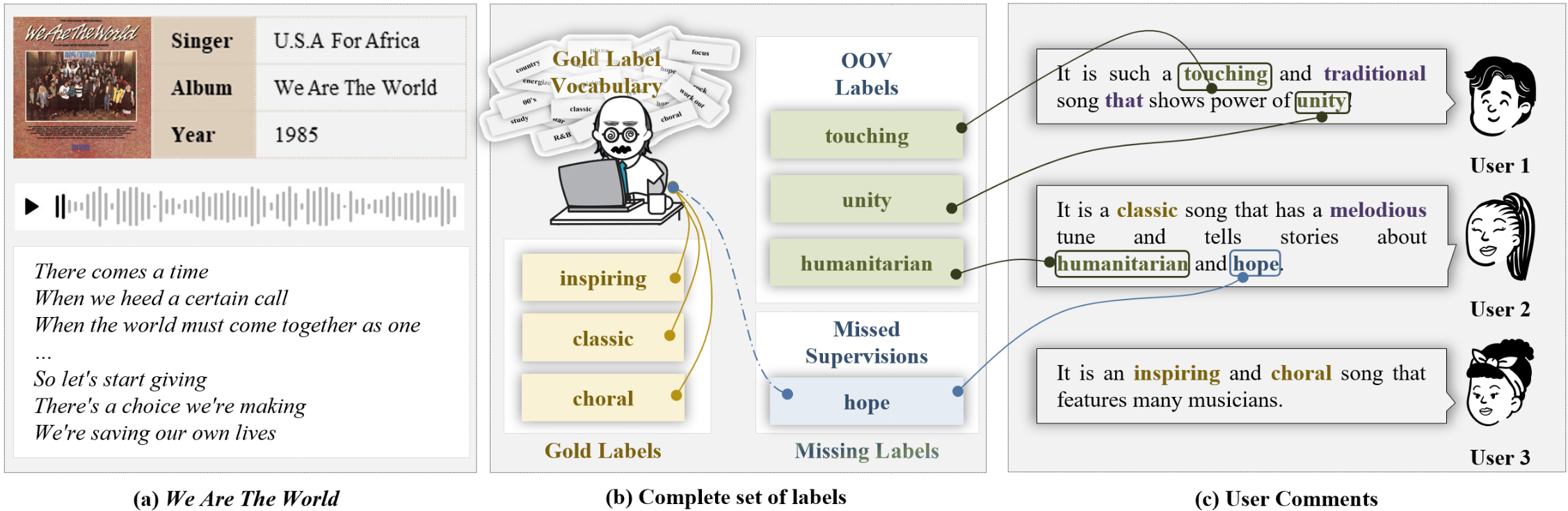}
    \caption{
    Illustration of (a) a song titled \textit{We Are The World}, (b) the complete set of labels for the song, and (c) the user comments on the song. The gold labels annotated by the expert only occupy part of all possible labels of the song. There lack of mappings to labels out of the gold label vocabulary~(OOV labels, e.g., \plabel{touching}) and some labels in the gold label vocabulary~(missed supervisions, e.g., \newsuper{hope}). Fortunately, all possible labels for the song are already presented in the user comments.
    }
    \label{fig:demo}
\end{figure*}
Towards sufficient music searching, especially at the industrial scale, an ideal solution is to hire experts to annotate the complete set of all possible labels for a song. As shown in \rfig{fig:demo}, an expert is hired to select the gold labels for a song~(i.e., \textit{We Are The World}) from the gold label vocabulary. However, such gold labels only take part of all possible labels of the song. The information missed by the gold labels exists in two scenarios. First, there lack of mappings to labels out of the gold label vocabulary~(OOV labels), e.g., \plabel{unity}. This is because we can never provide the annotator with a complete vocabulary covering all music characteristics and user searching preferences~\cite{simig2022open}. Second, even with a complete vocabulary, the mappings to some labels may also be missed by the expert~(missed supervisions), e.g., \newsuper{hope}. This is because the expert can never be serious and careful enough to pinpoint all the right labels from thousands or more candidate labels in the vocabulary~\cite{xiong2022extreme}. Therefore, to form the complete set of labels for a song, it is needed to obtain more diverse mappings to OOV labels and missed supervisions.\par
However, current multi-label classifiers~\cite{liang2018jtav,liang2020pirhdy}, including recent extreme multi-label learning methods~\cite{tagami2017annexml,wei2022survey}, fail to resolve the above challenge. Because they target to approach a vocabulary-size vector, where the elements indicating gold labels are set {to} 1 and other elements are set {to} 0. A few studies try to obtain more mappings using generative models~\cite{simig2022open}, self-training methods~\cite{zou2019confidence,xie2020self,xiong2022extreme}, etc. However, such new mappings are not diverse enough to make up for the missing information. Because these newly mapped labels are either semantically similar to the gold labels or are copied from instances with similar features. To our best knowledge, {it remains an open question how to automatically obtain the complete set of labels for instance given a small number of gold labels}.\par
Fortunately, there is a good chance to solve this question with the help of users. A helpful observation is that users themselves are ``experts'' in music searching. Specifically, the vocabulary the users used to comment on a song naturally matches the vocabulary they used to search for the song~\cite{gao2021advances}.
Besides, the user comments are continuously updated along with the changing of user's feelings and the latest Internet buzzwords, so they can reflect the dynamic {changes} in user searching preferences~\cite{mihret-atinaf-2019-sentiment}. 
As such, the complete set of OOV labels, missed supervisions, and gold labels for a song may already be presented in the user comments of the song, cf., \rfig{fig:demo} (c). This observation inspires us to harvest more labels from user comments. We believe such ``user-generated'' labels can largely enrich the expert-generated gold labels and support more sufficient music searching. Specifically, towards sufficient music searching, the newly generated labels should have two basic advantages. First, they should be adequately \textbf{valid} to serve as a label~(e.g., {\plabel{touching}} vs. \badlabel{that}) and to benefit music searching~({e.g., \plabel{touching}} vs. \badlabel{melodious}). Second, they should be \textbf{diverse} enough to semantically distinguish from the gold ones~(e.g., {\gold{classic}} vs. \badlabel{traditional}). This encourages us to study automated music labeling in a vital but under-explored setting --- given the limited gold labels of a song, the model is required to harvest more diverse and valid labels from the user comments for the song.\par
To this end, we propose \name, which is an iterative framework for harvesting more \underline{Di}verse and \underline{Va}lid labels for music. It consists of two key components, i.e., a binary classifier and a joint score function. The binary classifier takes the concatenation of user comments and a candidate label as input and outputs a confidence score indicating how likely the candidate label is a suitable label for the song. In the beginning, the binary classifier, which has been trained on the gold labels, has trouble predicting more diverse mappings than the gold ones. Inspired by \citet{xie2020self}, we use the pre-trained classifier to infer the pseudo-mappings to missed supervisions. In addition, 
we design a novel joint score function, which comprehensively measures the diversity and validness of a candidate label to infer the pseudo-mappings to OVV labels. In this way, the binary classifier can be fine-tuned on more diverse and valid labels than the gold ones. We repeat this process several iterations until the classifier can infer a complete set of labels for a song, namely, the classifier can barely infer any new pseudo-labels. We further collect a corpus\footnote{The corpus and our codes are available at \url{https://github.com/jingyaolliu/DiVa}.} from a Chinese online music platform\footnote{For convenience and saving spaces, all presented examples in this paper have been translated into English.}. It consists of {16,372} songs, each of which consists of user comments and gold labels. Towards data-driven evaluation, we also develop a testing set of label {1,000} songs, which have been densely annotated by experts based on the user comments. Experiment results show that, compared with state-of-the-art methods, the \name framework is more powerful in harvesting more diverse and valid labels for music than the state-of-the-art methods.  We hope our work can provide insights into automated music labeling and advance the research on finding the complete set of labels in real-world applications. 
In summary, we highlight our contributions as follows.
\begin{itemize}
    \item {We call attention to the information missed by the gold labels and study the automated music labeling in a challenging setting, where the model is required to produce more diverse and valid labels based on a small number of gold labels.}
    \item We propose an iterative framework~(\name) to harvest more diverse and valid labels from user comments. It makes a classifier infer more diverse and valid labels for music gradually with the help of a novel joint score function.
    \item The experiment reveals the superiority of \name in producing more diverse and valid mappings to OOV labels and missed supervisions as well as more complete sets of labels for songs.
\end{itemize}

\section{Problem Formulation}
\label{sec:pro}
\begin{figure}[t]
    \centering
    \includegraphics[width=0.7\linewidth]{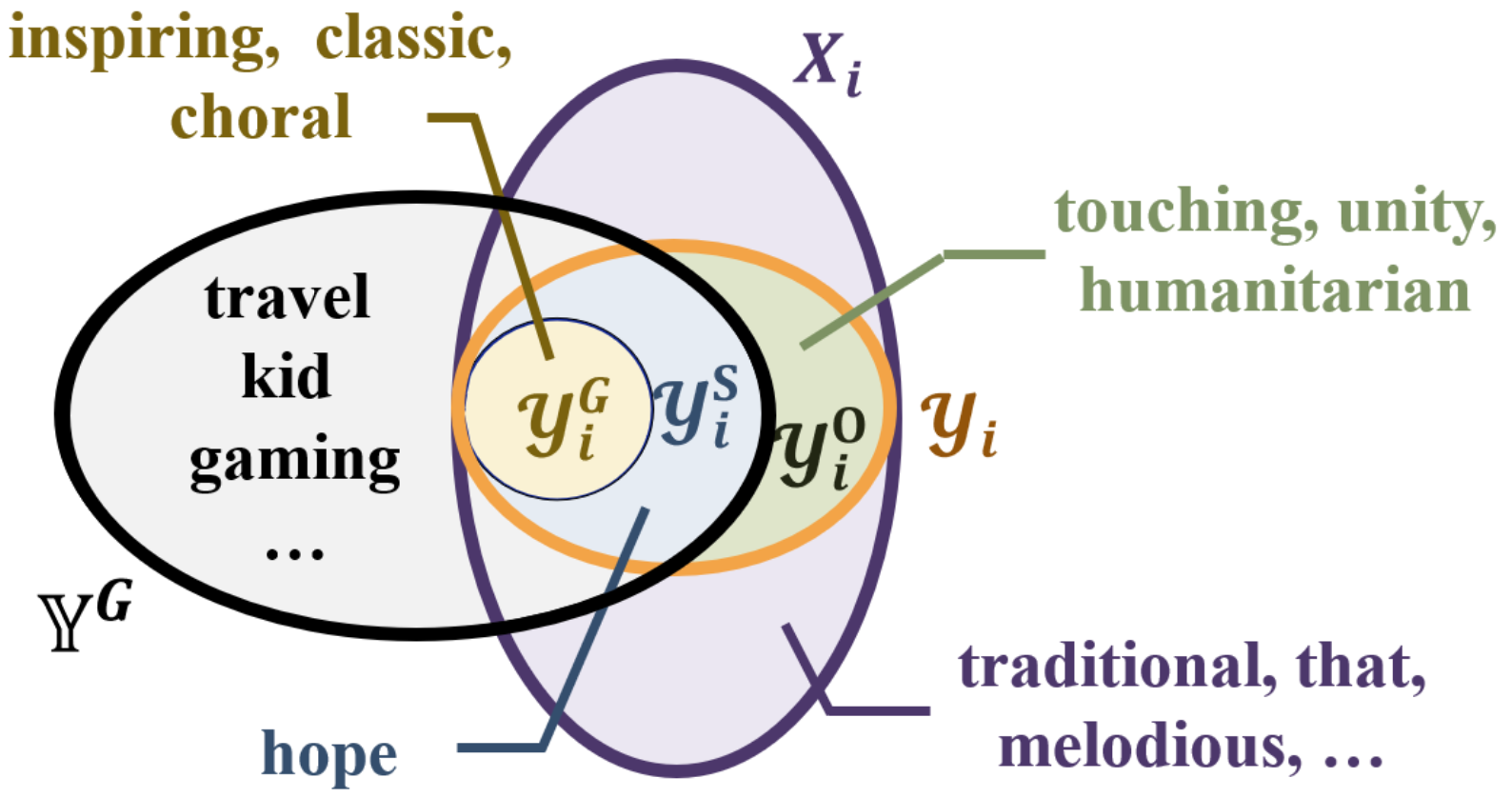}
    \caption{ Venn diagram for the labels of the $i^{th}$ song and the words in the user comments of the song
     }
    \label{fig:set}
\end{figure}

{Let $\{d_i, {X}_i, \mathcal{Y}_i^G, \mathcal{Y}_i\}_{i=1}^N$ denote a dataset of $N$ songs, where $d_i$, ${X}_i$, $\mathcal{Y}_i^G$, and $\mathcal{Y}_i$ are the user comments, the set of words tokenized from user comments,} gold labels, and the complete set of labels for the $i^{th}$ song, respectively. We assume $\mathcal{Y}_i$ is a proper superset of $\mathcal{Y}_i^G$ and a proper subset of ${X}_i$, cf., \rfig{fig:set}. Besides, $\mathcal{Y}_i$ is unknown both at the training and testing phases. Let $\mathbb{Y}^G=\bigcup_{i=1}^N \mathcal{Y}_i^G$ denote the gold label vocabulary. The OOV labels in \rfig{fig:demo} can be express as $\mathcal{Y}_i^O=\mathcal{Y}_i-\mathbb{Y}^G$ and the missed supervisions can be expressed as $\mathcal{Y}_i^S=\mathcal{Y}_i\cap \mathbb{Y}^G-\mathcal{Y}_i^G$. The task of harvesting more labels from the user comments is to find the ${\mathcal{Y}}_i^+ \subset {X}_i$ satisfying the following conditions:
\begin{equation}
\label{eq:obj}
    \mathop{\arg\max}\limits_{\mathcal{Y}_i^+} \frac{|\mathcal{Y}_i\cap(\mathcal{Y}_i^G\cup\mathcal{Y}_i^+)|}{|\mathcal{Y}_i\cup(\mathcal{Y}_i^G\cup\mathcal{Y}_i^+)|}.
\end{equation}
In other words, ${\mathcal{Y}}_i^+$ is required to cover as much as information missed by the gold labels~(i.e., $\mathcal{Y}_i^O \cup \mathcal{Y}_i^S$) using as few as labels. Note that, semantically similar labels~(e.g., \gold{classic} and \badlabel{traditional}) are treated as the same label in this task.
\begin{figure*}[t]
    \centering
    \includegraphics[width=0.88\textwidth]{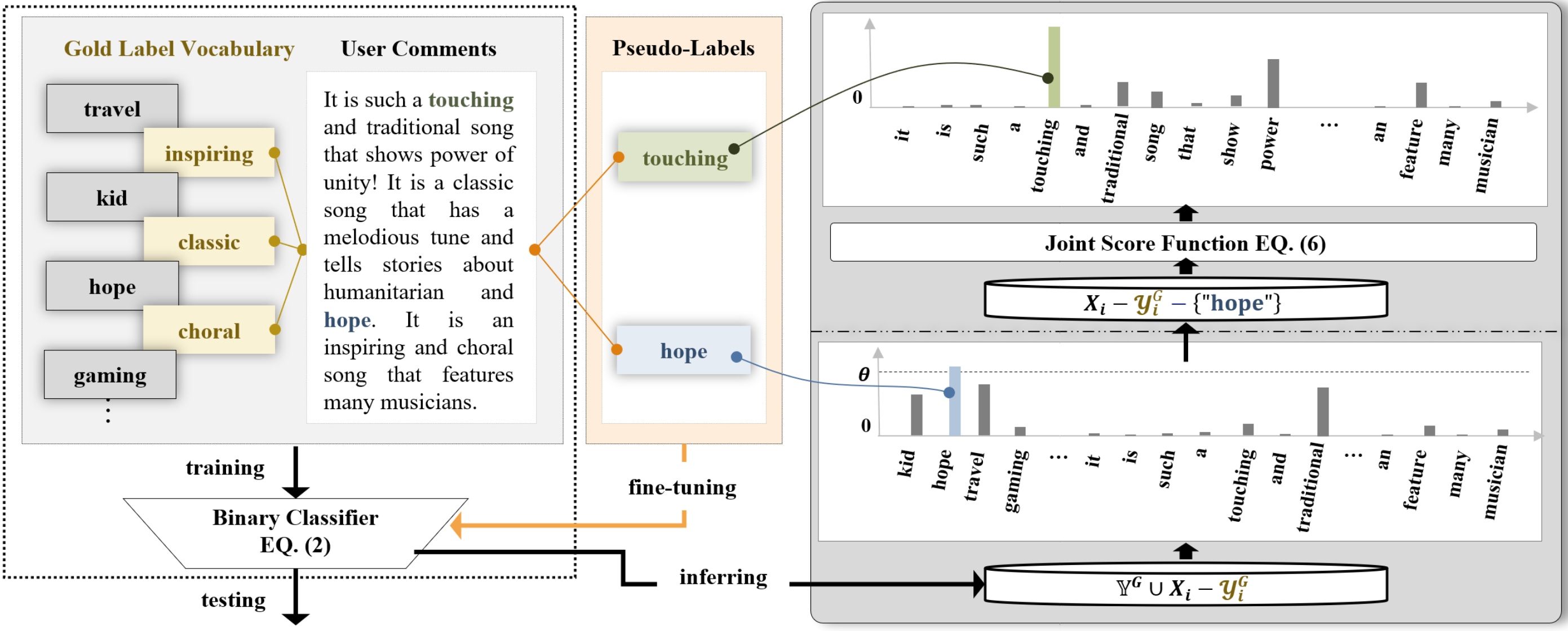}
    \caption{The proposed iterative framework~(\name). It consists of two key components, i.e., a binary classifier and a joint score function. The binary classifier is first trained on user comments and gold labels and then fine-tuned on pseudo-labels inferred from the pre-trained classifier and a novel joint score. This process continues until the classifier can barely produce any new pseudo-labels. During testing, the labels for a song are directly inferred from user comments via the optimized classifier.}
    \label{fig:diva}
\end{figure*}

\section{The DiVa Framework}
As shown in \rfig{fig:diva}, the \name framework starts from a binary classifier trained on paired user comments~($d_i$) and gold labels~($\mathcal{Y}_i^G$). The classifier is then used to infer the possibility distribution on candidate labels and select those with high confidence scores as pseudo-labels. Further, we design a joint score function to infer the possible distribution of the remaining candidates. All pseudo-labels are used to fine-tune the previous classifier. This process will repeat several iterations until the classifier can barely infer any new pseudo-labels or reach the maximum iterations. In the testing phase, we can harvest labels from user comments directly from the optimized binary classifier.

\subsection{Binary Classifier}
The multi-label classifier is limited to a close set of labels as it is required to fit a fixed-size {multi-hot} vector. To overcome this limitation, we lay our foundation on a binary classifier, which is required to fit a scalar-value label. Specifically, we pair the user comments~($d_i$) and a candidate label~(denoted as $y_c$) of the $i^{th}$ song as a new instance. If the candidate label is a gold label to the $i^{th}$ song, we label the new instance as 1; otherwise, the new instance is labeled as 0. At this moment, the binary classifier is equivalent to the conventional multi-label classifier. \par
To make it different, we make two changes to the binary classifier. First, we change the candidate labels for the $i^{th}$ song from labels in the gold label vocabulary to the words from user comments plus its gold labels~($\mathcal{Y}_i^G$). As such, we enrich the label vocabulary without increasing the number of candidate labels for the $i^{th}$ song. 
{Second, to better correlate document-level user comments with word-level labels, 
we use a contextual language model (i.e., XLNet~\cite{yang2019xlnet}) to represent the user comments and use the improved static word embeddings (i.e., X2Static~\cite{gupta2021obtaining}) to represent the labels.
}
We then concatenate the representations of user comments~($\vec{d_i}$) and a candidate label~($\vec{y_c}$) and use the concatenations as the input of the binary classifier, which is defined as
\begin{equation}
\label{eq:ebc}
    \mathcal{F}(d_i,y_c)= sigmoid(\vec{d_i}\oplus\vec{y_c}),    
\end{equation}
where $\oplus$ is the concatenate operation. The output of the $\mathcal{F}(d_i,y_c)$ is a confidence score indicating how likely $y_c$ is a gold label for the $i^{th}$ song. The objective of the classifier is to minimize the sum of the binary cross entropy between these confidence scores and the ground truth values~(0 or 1).\par

\subsection{Binary Classifier Inference} We then utilize the pre-trained classifier to infer the possibility distribution on the candidate labels, which involves the remaining labels of the gold vocabulary~($\mathbb{Y}^G$) and the remaining words tokenized from the user comments~($X_i$). Specifically, the candidate labels for the $i^{th}$ song are defined as $\mathbb{Y}^G\cup{X_i}-\mathcal{Y}_i^G$. The candidates with high confidence scores
are picked out as pseudo-labels for the following fine-tuning. Note that, such pseudo-labels~(e.g., \newsuper{hope}) can barely contribute to OOV labels missed by the annotation. This is because the classifier always assigns pseudo-labels to an instance by borrowing gold labels from other instances with similar features. This also explains why the candidates not in the vocabulary often get very low scores. An exception is the OVV candidate label \badlabel{traditional}, which gets a relatively high score. This is because \badlabel{traditional} is semantically similar to the gold label \gold{classic}. As a result, though the inferred pseudo-labels contribute to missed supervision, they cannot virtually form a complete set of labels for a song.

\subsection{Joint Score Function} We believe the above limitation will be largely mitigated if we ``teach'' the classifier what has not been acquired by it before, namely, fine-tuning the classifier on more diverse labels that are unknown at the previous training phase. Specifically, given the remaining candidate labels~(e.g., $X_i-\mathcal{Y}_i^G-\{\text{\newsuper{hope}}\}$), we design a joint score function to measure the diversity~(statistical importance and semantic novelty) and validness~(practical value and discrimination ability) of a candidate label. This function computes a joint value of the following scores.
    \paragraph{Statistical importance score~($SI$)} Following the tradition of information searching, we use TF-IDF to measure the statistical importance score of a candidate label. The higher the $SI$ score, the more important the candidate label is to a song from the view of data distribution.
    
    \paragraph{Semantic novelty score~($SN$)} If only an expert says a given candidate label is semantically different from the existing ones, it may be subjective. However, if many other experts also agree with that, the candidate label is most likely to be a semantic novel label. The analogy to this, we conduct $m$ groups of $K$-means clustering on the labels to compute the semantic novelty score of a candidate label against the existing ones. In this way, we get $m$ ``experts'', each of which has its own taxonomy and opinions about the labels. The confidence of $y_c$ being a semantic novel label is defined as 
    \begin{equation}
    \label{eq:sn}
        SN=\frac{1}{2}*\sum_{i=1}^{m}\frac{{1-\min\big[\mathcal{D}(y_c, C_i^1), \mathcal{D}(y_c, C_i^2),..., \mathcal{D}(y_c, C_i^K)\big]}}{m},
    \end{equation}
    where $C_1^K$ means the $K^{th}$ cluster of the $i^{th}$ ``expert'' and $\mathcal{D}(y_c, C_i^K)$ calculates the cosine similarity between $y_c$ and the center of $C_i^K$. 
    \paragraph{Practical value score~($PV$)} Similar to $SN$, if all experts agree that a candidate label is not suitable to be used as a label, then we think this candidate label is not valid enough from the point of practical values. Here, we treat the songs as the ``experts'' and aim to penalize candidates with low practical values~(e.g., below a pre-defined threshold $\tau$). Specifically, the practical value score depends on the confidence scores of the candidate label predicted on all songs and is defined as 
    \begin{equation}
    \label{eq:pv}
        PV=\left\{          
        \begin{matrix}             
        1, & \sum_{i=1}^N\frac{\mathcal{F}(X_i, y_c)}{N}\ge \tau, \vspace{1ex} \\               
        0, & \sum_{i=1}^N\frac{\mathcal{F}(X_i, y_c)}{N}< \tau.        
        \end{matrix}         
        \right.         
    \end{equation}
    \paragraph{Discrimination ability score~($DA$)} We believe a valid label, even presented in many songs, should be able to distinguish different songs. As such, we want to penalize candidates with low discrimination ability. Specifically, we use the coefficient of variation to describe the discrimination ability of a candidate label and define the $DA$ score as 
        \begin{equation}
        \label{eq:da}
        DA=\left\{          
        \begin{matrix}             
        1, & \frac{\sigma ({y_c})}{\mu (y_c)}\ge \tau, \vspace{1ex} \\               
        0, & \frac{\sigma ({y_c})}{\mu (y_c)}< \tau,        
        \end{matrix}         
        \right.         
    \end{equation}
    where $\tau$ is the same threshold used in $PV$, $\sigma({y_c})$ and $\mu ({y_c})$ computes the standard deviation and mean of the number of ${y_c}$ appearing in songs, respectively.
\par

Intuitively, we define the joint score function of the candidate label $y_c$ based on the above scores as
\begin{equation}
\label{eq:joint}
    \mathcal{J}(y_c)=SI*SN*PV*DA.
\end{equation}
The joint score function describes a unified objective to harvest labels with high semantic novelty and semantic novelty and avoid those with low practical value and discrimination ability. The candidate labels that get high joint scores are then selected as pseudo-labels, 
which involve OOV labels~(e.g., \plabel{touching}) and a few new supervisions missed before. 
These pseudo-labels, together with those selected by the pre-trained classifier, are used to fine-tune the classifier.

\subsection{Iteration}
Towards fast and stable optimization, we adopt the negative sampling strategy at the training/fine-tuning phase of the binary classifier. To construct a negative sample for a given song, we randomly select a label from the candidate labels except for the gold and pseudo ones and pair it with the user comments on the song. Note that, even if a candidate label, which may be a suitable label for a song, is accidentally used as a negative sample in the current iteration of training, the bad effect can be ignored as it will be selected as a pseudo-label via the following inference and be used to fine-tune the classifier. Besides, to balance the importance of rare and frequent pseudo-labels, we further leverage the subsampling strategy~\cite{mikolov2013distributed}. \par


Besides reaching the maximum, the iterations stop when the classifier can barely infer any new pseudo-labels. A straightforward solution is to define a threshold~(e.g., 50) --- if the number of new labels inferred by the binary classifier is below the threshold, we stop the iterations. However, it is hard to find a perfect threshold for different datasets, for example, 50 may be too large for datasets with a small-sized label vocabulary and too small for datasets with a large-sized label vocabulary. As a remedy, we estimate the stopping condition by evaluating the classifier's ability to infer new labels on long-tail labels, e.g., the performance on the PSP and PSnDCG metrics~\cite{jain2016extreme}. This is reasonable as the tail labels hold a big part of the labels missed by the gold ones.


\begin{table*}[t]
    \centering
    \caption{The number of labels w.r.t., the original corpus, training set and two testing set }
    \scalebox{1.0}{
    \begin{tabular}{ccccccccccccc}
    \hline
    \textbf{Dataset}& {\textbf{label}} &\multicolumn{3}{c}{\textbf{labels / song}}&&\multicolumn{3}{c}{\textbf{user comments / song}}&&\multicolumn{3}{c}{\textbf{words / user comment}}\\
        \cline{3-5}\cline{7-9}\cline{11-13}
       \textbf{(Number of songs)} &{\textbf{vocabulary size}} &\textbf{Avg.} & \textbf{Min.} & \textbf{Max.} &&\textbf{Avg.} & \textbf{Min.} & \textbf{Max.}&&\textbf{Avg.} & \textbf{Min.} & \textbf{Max.}\\
        \hline
        \textbf{Corpus~(16,372)} & 820 &  5.82 &  1 & 31 && 38.83 & 6 & 60 && 44.05 & 9 & 335 \\
        \hline
        \textbf{Training~(15,372)} & 816 & 5.82 & 1 & 31 && 38.84 & 6 & 60 && 44.04 & 9 & 335  \\
        \hline
        \textbf{\texttt{test-1}~(1,000)}& 425 & 5.78 & 1 & 27 && 38.76 & 11 & 60 && 44.15 & 9 & 303 \\
        \hline 
        \textbf{\texttt{test-2}~(1,000)}& 3312 & 19.41& 4 & 44 && 38.76 & 11 & 60 && 44.15 & 9 & 303 \\
        \hline  
    \end{tabular}
    }
    
    \label{tab:corpus}
\end{table*}
\section{Experiment}
\subsection{Experiment Settings}
\label{sec:setting}
\paragraph{Dataset}
To perform a systematic data-driven study, we gather a corpus of {16,372} songs, each of which consists of user comments and gold labels summarized by experts from the information retrieval department of a Chinese online music platform. The statistics are shown in Table~\ref{tab:corpus}. We then randomly select a testing set of 1000 songs with more than 10 user comments. Three senior experts from the same department are hired to annotate the testing set. At each time, a worker will be shown a song with its metadata~(title, artist, etc), user comments, and candidate labels~(stopwords are removed in advance), he is required to select proper labels for the song from the candidates after carefully go through the metadata and read the user comments. If needed, he can also use his own knowledge and searching necessary information online. We give a training session to the workers to help them fully understand our annotation requirements. The annotation tasks are released to the workers via an in-house website. On average, it takes about 20 minutes for a worker to annotate a song. Each song is randomly assigned to two workers. For the $i^{th}$ song, we calculate the Kappa coefficient score between the different workers' annotations. If the score is higher than 75\%, all selected candidates together with the gold labels are used as the complete set of labels $\mathcal{Y}_i$. Otherwise, we let the two workers discuss their disagreements and come up with the agreed results. After the annotation, we get two testing sets --- one has gold labels for every song~(denoted as \texttt{test-1}) and the other one has the complete set of labels for every song~(denoted as \texttt{test-2}). On average, there are about 5.8 gold labels for every song and about 19.4 labels for a song in \texttt{test-2}.
\paragraph{Evaluation Metrics}
Following the tradition, we evaluate \name on the commonly used multi-label classification metrics, which involve precision~(P), recall~(R), F1, and nDCG, and extreme multi-label learning metrics, which involve normalized PSP and PSnDCG~\cite{jain2016extreme}. The PSP and PSnDCG scores are not applicable for \texttt{test-2} as we assume that each song in \texttt{test-2} has a complete set of labels. Inspired by \citet{tandon2020dataset} and \citet{simig2022open}, to eliminate the inaccurate mismatch caused by synonym labels, we also employ the soft matching metrics. Given the predicted labels~(denoted as $\mathcal{Y}_i^*=\{{y}_j^*\}_{j=1}^{|\mathcal{Y}_i|}$) and the gold labels~(denoted as $\mathcal{Y}_i^G=\{{y}_k\}_{k=1}^{|\mathcal{Y}_i^G|}$) for the $i^{th}$ song, the soft precision is defined as 
    \begin{equation}
    \label{eq:softp}
        SP=\sum_{j=1}^{|\mathcal{Y}_i^*|}\frac{\max\big[\mathcal{D}({y}_j^*, {y}_1),...,\mathcal{D}({y}_j^*, {y}_{|\mathcal{Y}_i^G|}) \big]}{|\mathcal{Y}_i^*|},
    \end{equation}
    where $\mathcal{D}({y}_j^*, {y}_1)$, with the same definition in EQ.~(\ref{eq:sn}), returns the cosine similarity between $\vec{{y}}_j^*$ and $\vec{{y}_1}$. 
    The soft recall is defined as 
        \begin{equation}
        \label{eq:softr}
                SR=\sum_{k=1}^{|\mathcal{Y}_i^G|}\frac{\max\big[\mathcal{D}({y}_k, {y}_1^*),...,\mathcal{D}({y}_k, {y}_{|\mathcal{Y}_i^*|}^*) \big]}{|\mathcal{Y}_i^G|}.
    \end{equation}
    Accordingly, the soft F1 score is defined as $SF1=\frac{2*SP*SR}{SP+SR}$. Note that, the soft matching metrics are not suitable for methods that have represented labels via their own models. In line with EQ.~(\ref{eq:obj}), we also want to measure whether the predicted label can sufficiently make up for the information missed by the gold labels. As such, we design a coverage score on \texttt{test-2}, which is defined as Jaccard similarity between predicted labels and the complete set of labels, i.e., $Coverage=\frac{1}{N}*\sum_{i=1}^N\frac{\mathcal{Y}_i^* \cap\mathcal{Y}_i}{\mathcal{Y}_i^* \cup\mathcal{Y}_i}$. The coverage score for the gold labels in \texttt{test-2} is 0.284. For all metrics, higher values indicate better performances. Particularly, we place emphasis on the F1/SF1 and coverage scores, which are directly related to our ultimate goal, i.e., the ability to obtain a complete set of labels for each song.

\paragraph{Baselines} Towards a thorough comparison, we compare \name with the following baseline methods.
\begin{itemize}
    \item TF-IDF, which identifies labels based on the statistic features of the candidate label. It also contributes to the statistical importance score~($SI$) in our novel joint score function defined in EQ.~(\ref{eq:joint}).
    \item Multi-label Classification~(MLC), which takes user comments as input and outputs a label vocabulary-sized vector.
    \item LightXML~\cite{jiang2021lightxml}, which is a tree-based extreme multi-label learning method that uses dynamic negative sampling to boost the model performance on massive labels.
    \item Our binary classifier, which is define by EQ.~(\ref{eq:ebc}).
    \item nnPU~\cite{kiryo2017positive}, which avoids biased noises in binary classification via a non-negative risk estimator. We deploy this strategy based on our binary classifier.
    \item GROOV~\cite{simig2022open}, which labels songs through a generative model.
    \item ChatGPT~\cite{ouyang2022training}, which can be instructed to annotate text data via prompts~\cite{10.1145/3543873.3587368}.
    \item Noisy Student Training~(NST)~\cite{xie2020self}, which trains a classifier iteratively using pseudo-labels predicted on unlabeled instances. We deploy the NST framework on both MLC and binary classifiers, denoted as NST w/ MLC and NST w/ (EQ.~\ref{eq:ebc}), respectively. 
    \item Two variants of \name: \name -static, which, before starting the iterations, combines the pseudo-labels inferred from the initialized binary classifier and the joint score function as labels; \name -light, which only uses the pseudo-labels inferred from the current iteration to fine-tune the binary classifier.
\end{itemize}
\begin{table*}[t]
    \centering
    \renewcommand\arraystretch{1.2}
    \caption{Performances of all methods on \texttt{test}-1/\texttt{test}-2 w.r.t., precision~(P), recall~(R), F1, nDCG, normalized PSP~\cite{jain2016extreme}, normalized PSnDCG~\cite{jain2016extreme}, soft precision~(SP), soft recall~(SR), soft F1~(SF1), and the coverage score. We mark the values indicating the best performance in bold and the values indicating the second-best performance with underlines.}
    \scalebox{0.8}{
    \begin{tabular}{lccccccccccccc}
    \hline\hline
        \multirow{2}{*}{\textbf{Method}} & \multicolumn{4}{c}{\textbf{MLC } } && \multicolumn{2}{c}{\textbf{XML }} & &\multicolumn{3}{c}{\textbf{Soft matching }} &&\multirow{2}{*}{\textbf{Coverage}}\\ 
        \cline{2-5}\cline{7-8}\cline{10-12}
         & \textbf{P} & \textbf{R} & \textbf{F1} & \textbf{nDCG} && \textbf{PSP} & \textbf{PSnDCG} & &\textbf{SP} & \textbf{SR} & \textbf{SF1}&&\\ 
        \hline
        \textbf{TF-IDF} & 0.015/0.082 & 0.090/0.346 & 0.021/0.119  &0.046/0.247 && 0.061/- & 0.057/- && 0.064/0.245 & 0.164/0.415 & 0.081/0.282 && 0.139 \\  \hline
        \textbf{MLC} & 0.044/0.063 & 0.013/0.007 & 0.018/0.012 & 0.044/0.062 && 0.152/- & 0.137/- && 0.050/0.071 & 0.019/0.012 & 0.025/0.020 && 0.283 \\
        \textbf{LightXML~\cite{jiang2021lightxml}} & 0.405/0.819 & 0.488/0.271 & 0.411/0.396 & 0.539/\underline{0.834} && 0.449/- & 0.418/- && \underline{0.441}/\underline{0.877} &  0.521/0.316 & 0.449/0.454 && 0.395 \\

        \textbf{EQ. (\ref{eq:ebc})} & \underline{0.443}/\underline{0.836} & 0.691/0.360 & \textbf{0.510}/0.492 & \textbf{0.686}/\textbf{0.858} && 0.653/- & \textbf{0.594}/- && \textbf{0.475}/\textbf{0.893} & 0.713/0.404 & \textbf{0.544}/0.546 && 0.420 \\
        \textbf{nnPU~\cite{kiryo2017positive} w/ EQ. (\ref{eq:ebc})} & 0.070/0.189 & \textbf{0.980}/\underline{0.737} & 0.128/0.298 & \underline{0.654}/0.690 && \textbf{0.931}/- & \underline{0.513}/- && 0.126/0.318 & \textbf{0.982}/\underline{0.766 }& 0.220/0.446 && 0.178 \\
        \hline
        \textbf{GROOV~\cite{simig2022open}} & \textbf{0.512}/\textbf{0.844} & 0.412/0.200 & 0.426/0.310 & 0.295/0.244 && 0.327/- & 0.203/- && - & - & - && 0.206
         \\   
        \textbf{ChatGPT~\cite{ouyang2022training}} & 0.128/0.270 & 0.141/0.098 & 0.110/0.129 & 0.130/0.177 && 0.105/- & 0.105/- && - & - & - && 0.239\\  \hline
        \textbf{NST~\cite{xie2020self} w/ MLC} & 0.096/0.155 & 0.022/0.013 & 0.034/0.024 & 0.098/0.155 && 0.189/-
        & 0.185/-
        && 0.105/0.184& 0.029/0.021 & 0.044/0.037 && 0.286\\  
        \textbf{NST~\cite{xie2020self} w/ EQ. (\ref{eq:ebc})} & 0.337/0.775 & 0.746/0.483 & \underline{0.444}/0.584 & 0.603/0.802 && 0.645/-
        & 0.462/-
        && 0.378/0.841 & 0.765/0.521 & \underline{0.487}/0.633 && 0.484 \\  
        \hline
        \textbf{\name -static} & 0.175/0.566 & \underline{0.897}/\textbf{0.811} & 0.281/\underline{0.657} & 0.462/0.703 && \underline{0.835}/-
        & 0.404/- && 0.222/0.707 & \underline{0.906}/\textbf{0.831} & 0.344/\textbf{0.757} &&{0.517} \\ 
        \textbf{\name -light}  & 0.231/{0.694} & 0.722/0.629 & {0.331}/0.648 & 0.564/0.753 && {0.644}/- 
        & {0.451}/- && 0.274/0.784 & {0.745}/0.663 & 0.383/0.709 && \underline{0.550}\\ 
        \textbf{\name} & 0.239/0.727 & 0.716/0.631 & 0.341/\textbf{0.665} & 0.569/0.782 && 0.643/- 
        & 0.444/- && 0.282/0.819 & 0.739/0.665 & 0.392/\underline{0.724} && \textbf{0.570}\\  \hline\hline
\end{tabular}
    }
    \label{tab:main-1}
\end{table*}
\paragraph{Implementation Details}
The \name framework and the neural baselines are trained/fine-tuned on 8 Nvidia A6000 GPUs using Adam optimizer with a learning rate=0.001. We use the pre-trained XLNet-base model provided by \citet{yang2019xlnet}) to represent the user comments. As for candidate labels, following~\citet{gupta2021obtaining}, we get the X2Static embeddings based on the XLNet-base model and CBOW model. All hyperparameters are tuned on the training set. Each baseline is optimized to gain its best performance.

\subsection{Results and Observations}
\label{sec:results}
As mentioned in Sec.~\ref{sec:pro}, our ultimate goal is to predict the complete set of labels for a song. Table~\ref{tab:main-1} demonstrates that the proposed \name framework is the most desired solution to the goal. Because it obtains superior results~(best F1 score, best coverage score, and second best SF1 score) on \texttt{test-2}, suggesting that \name, compared to other methods, can cover more diverse information using fewer labels. We believe this is achieved by the iterative training/fine-tuning of the gold labels and pseudo-labels inferred from the pre-trained binary classifier and the novel joint score function. We also make the following observations.
\begin{itemize}
    \item When evaluating on \texttt{test-1}, EQ.~(\ref{eq:ebc}) obtains better F1 and SF1 scores yet much worse PSP score than of \name -static. These results indicate that if we equip EQ.~(\ref{eq:ebc}) with the joint score function, it is more powerful to handle an incomplete set of labels. This assumption is further demonstrated by the evaluation on \texttt{test-2}, where \name -static has higher F1, SF1, and coverage scores than EQ.~(\ref{eq:ebc}).
    \item The \name -light method gets worse worse F1, SF1, coverage scores than \name. We conjecture this is because only using pseudo-labels to fine-tune the classifier makes the classifier overfit to pseudo-labels during the iterations. As a result, though the binary classifier learned from \name -light has the ability to produce new information missed by the gold labels, it loses the ability to produce gold labels. 
    \item From the comparison between NST~\cite{xie2020self} and \name, we notice that NST always works better than \name on \texttt{test-1} but works worse than \name on \texttt{test-2}. This is because the goal of self-training methods is to approach the mappings to gold labels and thus these methods, including NST, are good at handling missed supervision. Whereas, the \name framework forces the classifier to learn mappings to OOV labels as well as missed supervisions and thus is better at forming complete sets of labels for songs.
    \item Compared with \name and other methods, nnPu~\cite{kiryo2017positive} w/ EQ.~(\ref{eq:ebc}) gets the best PSP scores on \texttt{test-1}. Thus, it has a big chance to get a good performance on \texttt{test-2}. However, its coverage score is only slightly better than the worst one~(TF-IDF) and much worse than the best one~(\name). This is because the labels produced by nnPu contain too much noise. Another piece of evidence is that nnPu gets very high R and SR scores but rather low P and SP scores on \texttt{test-1} and \texttt{test-2}.
    \item We further observe that a better understanding of labels can greatly improve performance. Specifically, the methods using indexes to represent labels~(MLC and NST w/ MLC) always work worse than those using tree~(lightXML~\cite{jiang2021lightxml}) or embeddings to represent labels~(GROOV~\cite{simig2022open} and \name).
    \item To better instruct ChatGPT~\cite{ouyang2022training}, we have tried all off-the-shelf prompts that are designed for text annotation and utilize the prompt with the best performances~\cite{10.1145/3543873.3587368,moller2023prompt,gilardi2023chatgpt} to predict labels for songs from user comments. However, ChatGPT doesn't produce promising results as it has done in other tasks. We find a possible explanation in \citet{gupta2021obtaining} that contextual language models are not always welcome by word-level interpretability.  
    \item Compared to conventional MLC, our binary classifier defined by EQ.~(\ref{eq:ebc}) performs well not only on \texttt{test-1} but also \texttt{test-2}. This is because instead of using the labels in the gold label vocabulary, we use words from use comments as candidate labels, the number of which is less than the gold vocabulary size. So it can learn more reliable mappings in a lower-dimensional space. Besides, we use X2Static~\cite{gupta2021obtaining} to encode labels into vectors. So the labels have a better correlation with each other and user comments. This is also why EQ.~(\ref{eq:ebc}), as a supervised method, can produce some labels beyond the gold label vocabulary.
\end{itemize}
\begin{figure}[t]
    \centering
    \includegraphics[width=\linewidth]{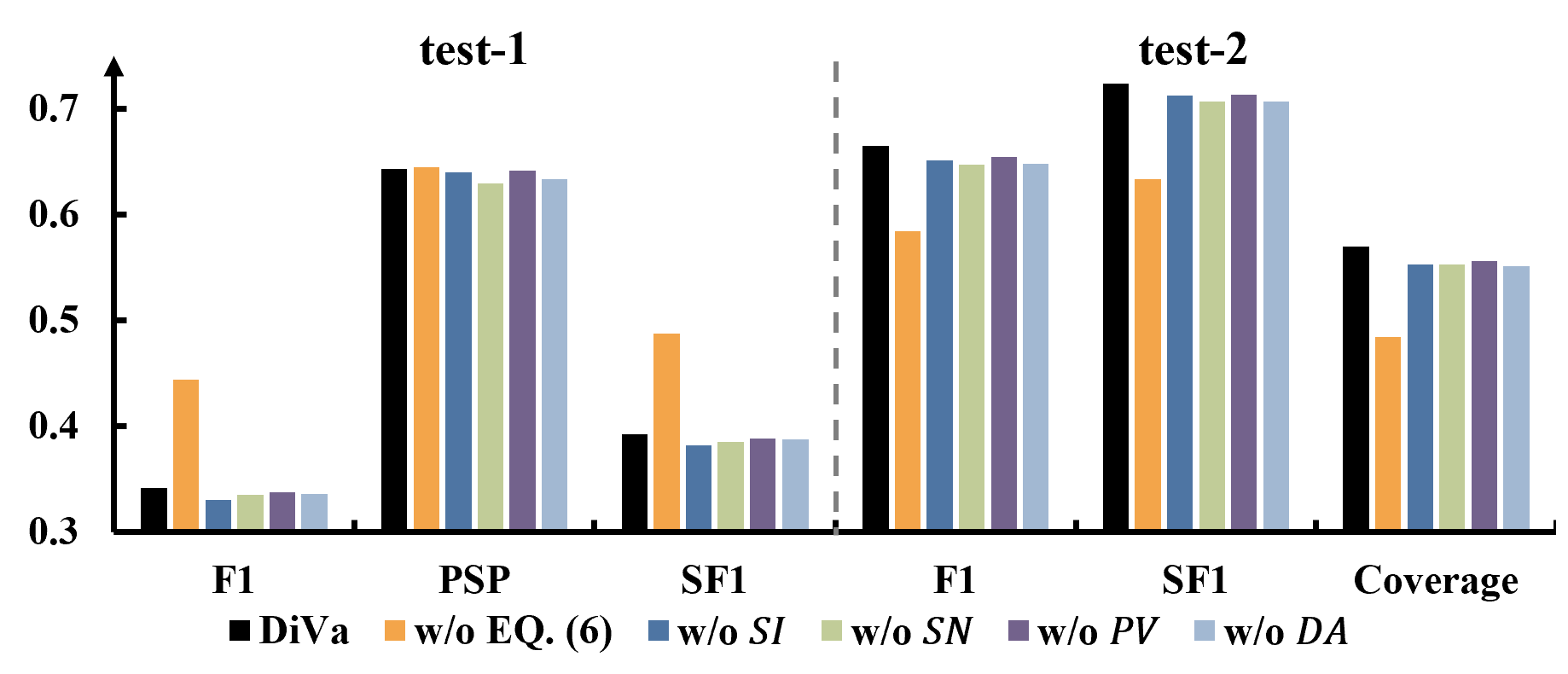}
    \caption{Performances w.r.t., different scores of EQ.~(\ref{eq:joint}) on \texttt{test-1}~(left) and \texttt{test-2}~(right)}
    \label{fig:ab}
\end{figure}
\begin{figure*}[htpb]
    \centering
    \includegraphics[width=0.7\linewidth]{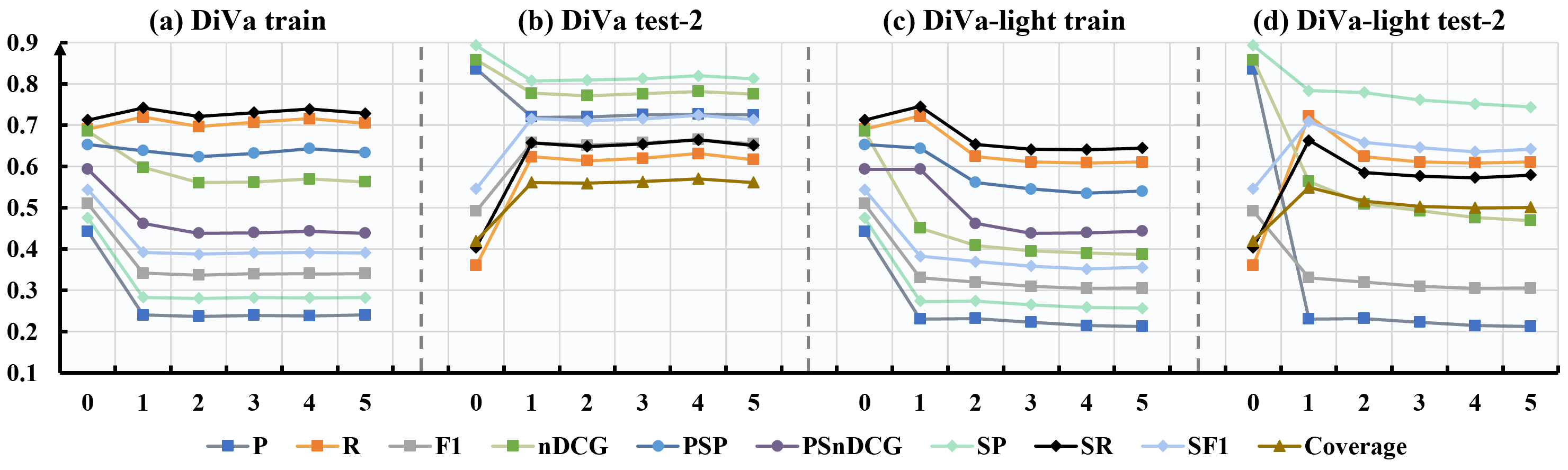}
    \caption{Performances of (a) \name on the training set, (b) \name on \texttt{test-2}, (c) \name -light on the training set, and (d) \name -light on \texttt{test2} over iterations.}
    \label{fig:lt}

\end{figure*}
\subsection{Ablation Study on different scores of EQ.~(\ref{eq:joint})}
As shown in \rfig{fig:ab}, we perform an ablation study to investigate the effect of the novel joint score function~(EQ.~(\ref{eq:joint})) and the effect of $SI$, $SN$, $PV$, $DA$ scores in measuring a candidate label. Note that, if we remove the novel joint score function from \name during the iterations, we will get NST~\cite{xie2020self} w/o EQ.~(\ref{eq:joint}). \rfig{fig:ab} presents the performances with different joint score functions on \texttt{test-1} and \texttt{test-2}, respectively. From the comparisons between \name against \name w/o $SI$, w/o $SN$, w/o $PV$, and w/o $DA$, we can see that any one of the scores has little impact on \name. However, when these scores are assembled in the joint score function, there presents a strong influence on \name. Specifically, we see a big rise~(about 0.1) of F1 and SF1 scores when comparing \name and \name w/o EQ.~(\ref{eq:joint}) on \texttt{test-1}. Conversely, we see a big drop~(about 0.08) of all metrics when comparing two methods on \texttt{test-2}. This is because EQ.~(\ref{eq:joint}) is designed to find more diverse and valid pseudo-labels beyond gold labels to fine-tune the binary classifier. Thus, these pseudo-labels are ``noises'' for the ground truth values in \texttt{test-1} but the right answers for the ground truth values in \texttt{test-2}, resulting in the rise and drop in \rfig{fig:ab}. 
    

\subsection{Performance over iterations}
We set the iteration number of \name and \name -light based on their results, mainly the PSP and PSnDCG scores, on the training set. Note that, the classifier at the $0^{th}$ iteration is only trained on gold labels. As we can see from \rfig{fig:lt}~(a) and \rfig{fig:lt}~(c), the best iteration number for \name is 4 and that for \name -light is 1.  For example, the results of \name shown in Table~\ref{tab:main-1} are obtained from the binary classifier fine-tuned at the $4^{th}$ iteration. For auxiliary analysis, we also report the results of \name and \name -light on \texttt{test-2} produced by the binary classifier optimized in every iteration in \rfig{fig:lt}~(b) and  \rfig{fig:lt}~(d). In keeping with the training set, \name also gets its best performance at the $4^{th}$ iteration. We observe a trend that the precision scores are becoming smaller gradually with the increase of iterations on \texttt{test-1} and \texttt{test-1}. This is because, at each iteration, the classifier is fine-tuned on more pseudo-labels, many of which are regarded as ``noises'' by the gold ones. These ``noises'' has a negative effect on the precision scores. However, the decrease presented in \rfig{fig:lt}~(c) is much sharper than that  presented in \rfig{fig:lt}~(a). This is because, as mentioned in Sec.~\ref{sec:results}, the classifier, only fine-tuned on pseudo-labels, is overfitting to the pseudo-labels and loses the ability to predict gold labels. Accordingly, it cannot get good performances on \texttt{test-2}, as shown in \rfig{fig:lt}~(d).

\subsection{Case Study}
\label{sec:case}
\begin{table*}[t]
    \centering
     \caption{Examples of two songs with \tgold{gold labels}, \tfull{complete set of labels} and labels produced by different methods. }
     \scalebox{0.8}{
    \begin{tabular}{lp{15cm}}
    \hline\hline
    \textbf{Title: \textit{The Last of The Mohicans}}    & \textbf{Artist: Alexandro Querevalú}\\
    \hline
     \textbf{Gold labels} & \tgold{soul-touching, aesthetic, trouble, passionate, sad} \\
     \hline
     \textbf{Complete set of labels} & \tfull{melodious, proud, helpless, life, era, soul-touching, male, surging, America, shot, future, world, moving, the sound of nature, turbulent, passionate, sad, nation, lonely, touching, story, aesthetic, trouble, slave, war}\\
   \hline
     \textbf{LightXML~\cite{jiang2021lightxml}}    & \tgold{aesthetic}, \tgold{passionate}, single, \tfull{touching}, \tfull{lonely}, \tfull{moving}, \tfull{world}  \\
     \hline
     \textbf{EQ.~(\ref{eq:ebc})}& \tgold{passionate}, \tgold{sad}, single, \tfull{touching}, \tfull{story}, \tfull{lonely}, \tfull{moving}, \tfull{world}, \tfull{soul-touching}, \tgold{aesthetic} \\
     \hline
     \textbf{nnPU~\cite{kiryo2017positive}  w/ EQ.~\ref{eq:ebc}}  & \tgold{passionate},  \tfull{war}, \tfull{proud}, \tgold{sad}, single, \tfull{life}, \tgold{story}, cry, \tfull{moving}, \tfull{world}, \tgold{future}, video, \tfull{era}, \tfull{touching}, witness, peace, eat, pure, hope, \tfull{the sound of nature}, \tnew{scared}, profit, street, \tgold{trouble}, remember, poor, the Orient \\
     \hline
     \textbf{GROOV~\cite{simig2022open}}    & healing, peaceful, memory, \tnew{classic}, love song, love, affectionate, single, \tfull{story}, galaxy, youth, sunshine, self-improvement, enthusiasm, \tgold{aesthetics}, \tnew{light music}, adoration, \tnew{bgm}, \tgold{soul-touching}, archaic rhyme \\
     \hline
     \textbf{ChatGPT~\cite{ouyang2022training}}   & \tnew{original ecological music}, \tnew{folk music}, humanistic concern, historical significance, reflection on human civilization, cultural diversity, \tnew{tragic}, \tnew{struggle and frustration} \\
     \hline
     \textbf{NST~\cite{xie2020self} w/ EQ.~(\ref{eq:ebc})}   & \tfull{world}, \tfull{story}, single, \tgold{sad}, \tgold{passionate}, \tfull{moving}, \tfull{touching}, \tfull{lonely}, \tgold{aesthetic} \\
     \hline
     \textbf{\name}    & \tfull{moving}, \tgold{sad},  single, \tfull{touching}, \tfull{story}, \tfull{lonely}, \tgold{passionate}, \tgold{aesthetic}, \tfull{turbulent}, \tfull{world}, \tfull{male}, \tfull{slave}, Spanish, \tnew{play wind instruments}, \tfull{the sound of nature}, \tfull{nation}, \tfull{era}, \tnew{soul-stirring}, \tnew{shock}, good faith, \tgold{sad}, \tfull{war}, democratic, \tgold{soul-touching}, \tfull{life}, \tfull{surging}, \tfull{melodious} \\
     \hline
     \hline
        \textbf{Title: \textit{Mermaid}}    & \textbf{Artist: Skott}\\
     \hline
     \textbf{Gold labels}    & \tgold{meet}\\
     \hline
     \textbf{Complete set of labels}    & \tfull{deep love, spacious, peaceful, chilly, appeal, thunder, clip, forest, bewitching, ethereal, meet, beautiful, happy, affectionate, pirate, story, stunning, love} \\
     \hline
     \textbf{LightXML~\cite{jiang2021lightxml}}    & \tfull{affectionate}, \tfull{deep love}, \tfull{beautiful}\\
     \hline
     \textbf{EQ.~(\ref{eq:ebc})}& \tfull{love}, \tfull{deep love}, \tfull{affectionate}, \tfull{story}, \tfull{beautiful}\\
     \hline
     \textbf{nnPU~\cite{kiryo2017positive}  w/ EQ.~\ref{eq:ebc}}  & \tfull{affectionate}, \tfull{deep love, story}, video, \tfull{beautiful}, \tfull{stunning}, share, \tfull{peaceful}, \tfull{forest}, \tfull{ethereal}, hope, \tfull{happy}, vocal cords, style, worth, anticipation, tone colour, \tgold{meet, clean}, treasure, wish, collect, human heart\\
     \hline
     \textbf{GROOV~\cite{simig2022open}}    & healing, galaxy, memory, \tfull{peaceful}, love song, \tfull{affectionate}, \tfull{love}, sad, classic, youth, sunshine, \tgold{meet}, single, bgm, rhythm, passionate, soul-touching\\
     \hline
     \textbf{ChatGPT~\cite{ouyang2022training}}& \tnew{famous singer Thunder sister}, \tnew{mermaid}, sea monster, \tnew{Skott}, Bjork, Huawei, Edited video of Ice and Snow 2\\
     \hline
     \textbf{NST~\cite{xie2020self} w/ EQ.~(\ref{eq:ebc})}   &  \tfull{story}, \tfull{love}, \tfull{deep love}, \tfull{affectionate}, \tfull{ethereal}, \tfull{forest}, \tfull{beautiful},  \tfull{happy}, \tfull{peaceful} \\
     \hline
     \textbf{\name}    & \tfull{beautiful}, \tfull{love}, \tfull{story}, \tfull{affectionate}, \tfull{deep love}, \tfull{pirate, ethereal}, \tfull{forest, bewitching}, \tfull{spacious}, \tfull{chilly}, on the road, \tfull{peaceful, appeal}, \tfull{happy}, \tfull{stunning}, \tgold{meet}\\
     \hline\hline
    \end{tabular}
}
    \label{tab:case}
\end{table*}
We present two examples of two songs~(\textit{The Last Of The Mohicans} and \textit{Mermaid}) with {gold labels}, {complete set of labels} and labels produced by different methods in Table~\ref{tab:case}. 
Among all methods, our \name framework can cover most labels of the complete set with the least bad labels. 
As explained in Sec.~\ref{sec:results}, the supervised methods~(lightXML, our binary classifier EQ.~(\ref{eq:ebc}), and NST w/ EQ.~(\ref{eq:ebc})) have the ability to infer missed supervision because of their extra attentions on labels. Surprisingly, although nnPU w/ EQ.~(\ref{eq:ebc}) and GROOV have produced many labels, most of them are bad ones. We also have an interesting observation that, ChatGPT, even with carefully designed prompts, tends to model the automated music labeling as a summarization task and tends to produce phrases rather than words.


\section{Related Work}
\paragraph{Extreme Multi-label Learning} 
The widely used XML methods can be divided into three types~\cite{liu2021emerging,wei2022survey}: the embedding-based methods try to map both the features and the labels into a joint low-dimensional space~\cite{tagami2017annexml,guo2019breaking}, the tree-based methods partition labels or instances into a tree~\cite{you2019attentionxml,chang2020taming,jiang2021lightxml}, and one-vs-all methods learns a binary classifier for each label separately~\cite{babbar2013flat,zhang2018binary}. Although such XML methods can learn more reliable mappings from massive labels, they require full label coverage and full supervision to train the classifiers. Note that the binary classifier used in \name differs from the one-vs-all XML methods in two aspects. First, instead of learning a classifier for each label,  we only design one binary classifier for all the concatenation of user comments and candidate labels. This largely reduces the number of parameters that need to be learned. Second, instead of using the full label set as candidates, we only use the gold labels and the words from user comments as the candidate labels of the song in the training phase. This largely reduces the number of candidates that need to be estimated.
\paragraph{Open Vocabulary Classification} \citet{simig2022open} introduces the open vocabulary XML classification for the first time. It also proposes GROOV, a generative model that can tag textual context with a set of labels from an open vocabulary. Although the GROOV model can enlarge the number of labels, the newly generated labels share a lot of similarities with the gold ones in the semantic space.
Another work on open vocabulary classification is \citet{xiong2022extreme}, which is also the only work, as far as we know, that concerns both incomplete label coverage and incomplete supervision. However, this approach is more suitable for instances with a small number of labels and can hardly apply to automated music labeling. Because it trains the model on too many noise pseudo labels that could lead the user searching to wrong songs with big chances. It also has a strict assumption that all instances must have label-like~(e.g., title) phrases. This is not always satisfied in real-world scenarios. 
\paragraph{Self-training} The self-training methods work by training a classifier iteratively by assigning pseudo-labels to unlabeled instances~\cite{zou2019confidence,van2020survey,amini2022self}. For example, Noisy Student Training~\cite{xie2020self} has three steps: train a teacher classifier on labelled instances, predict the pseudo-labels on unlabeled instances, and train a student classifier on both labelled and pseudo-labelled data. However, the self-training classifier tends to assign similar labels to instances with similar features. This is not helpful to capture the complete picture of user searching preferences on a single song. 
\paragraph{Positive-Unlabeled~(PU) Learning} The incomplete supervision issue can be also found in the PU learning setting, where only some of the positive instances are labeled~\cite{bekker2020learning}. One of the most popular PU learning methods is nnPU~\cite{kiryo2017positive}, which treats the unlabeled instances as negative instances and weights the unlabeled data via a non-negative risk estimator to avoid biased noises. Generally, such methods are designed for binary classification settings. In this paper, we can change the multi-label classification task to a binary classification task by pairing the user comments and a candidate label as an instance. This makes it possible to apply the PU learning methods in multi-label classification settings. However, this doesn't mean that such methods are suitable for the proposed task. Because these methods aim to get better performances on the seen supervision instead of digging the unseen supervision. Besides, they require the prior class distribution to optimize the classifier.

\paragraph{Active learning} The proposed iterative framework~(\name) is also slightly relevant to active learning, which selects instances for annotating and training in an interactive process~\cite{settles1995active,ren2021survey}. However, instead of reducing the cost during the annotation stage, our work aims to make the best of the labelled data after the annotation stage. Besides, instead of hiring experts for annotation, we obtain pseudo-labels from a joint score function.

\section{Conclusion}
In this paper, we call attention to the information missed by the gold labels and highlight the importance of the complete set of labels for a song towards sufficient music searching. After a detailed investigation of the missing labels, user comments, and current studies, we'd like to take a step forward and study automated music labeling in an essential yet under-exploring setting --- given limited gold labels, the model is required to harvest more diverse and valid labels from the user comments to form a complete set of labels for the song. Further, we develop an iterative framework~(\name) to learn a desired classifier for the proposed automated music labeling task. We also design a novel joint score function that can offer more diverse and valid pseudo-labels to fine-tune the classifier. Experiments on two testing sets reveal the superiority of \name in covering as much information with as few labels. Besides, the comparison with other iterative methods demonstrates the effectiveness of the novel joint score function in harvesting more diverse and valid labels. We hope our work can inspire future research on automated music labeling and shed light on multi-label learning with missing information in real-world scenarios.

\section*{Acknowledgements}
This work was supported in part by the National Natural Science Foundation of China (No. 62206191 and No. 62272330);
 in part by the China Postdoctoral Science Foundation (No.2021TQ0222 and No. 2021M700094); 
in part by the Natural Science Foundation of Sichuan (No. 2023NSFSC0473), 
and in part by the Fundamental Research Funds for the Central Universities (No. 2023SCU12089 and  No. YJ202219).
\clearpage
\bibliographystyle{ACM-Reference-Format}
\bibliography{main.bbl}
\clearpage
\appendix
\label{sec:appendix}

\section{Implementation Details}
\subsection{Baselines}
\paragraph{TF-IDF}
We deploy the classic TF-IDF, which is defined as 
\begin{equation}
    \text{TF-IDF}(y_c, d_i) = \frac{u_{y_c}}{\sum\limits_{d_i} {u_{y_c}}} \times \log\frac{N}{|\{d_j: y_c \in d_j\}_{j=1}^N|},
\end{equation}
$u_{y_c}$ and $\sum\limits_{d_i} {u_{y_c}}$ denote the number of occurrences of $y_c$ in the current user song comments $d_i$ and the total number of occurrences of all words tokenized from the comments, respectively. The left side of $*$ indicates the relative frequency of the candidate label $y_c$ within the current song comments, and the right side of $*$ is related to the number of songs with $y_c$ in their user comments.
\paragraph{Multi-label Classification~(MLC)} We use the user comments of a song to predict a vector, whose dimension is equal to the size of gold label vocabulary. Each element of the vector indicates the probability of $y_c$ being an accurate label for the $i^{th}$ song. The multi-label classifier is formulated as 
\begin{equation}
    F(d_i,\mathbb{Y}^G)=sigmoid(\vec{d_i}),
\end{equation}
where $\vec{d_i}$ is encoded from XLNet~\cite{yang2019xlnet}. The objective of the classifier is to minimize the sum of the binary cross entropy between the predicted and ground truth vectors. 
\paragraph{LightXML~\cite{jiang2021lightxml}} We use the codes released by the original paper \url{https://github.com/kongds/LightXML}.
\paragraph{nnPU~\cite{kiryo2017positive} w/ EQ.~(\ref{eq:ebc})} The original loss function of EQ.~(\ref{eq:ebc} is defined by the binary cross entropy, i.e.,

\begin{equation}
\begin{aligned}
    \mathcal{L}=&\sum_{i=1}^{N}\sum\limits_{y_c}BCE(y^*, \mathcal{F}({d_i,y_c}))\\
    =&-\sum_{i=1}^{N}\sum\limits_{y_c}y^*log(\mathcal{F}({d_i,y_c}))+(1-y^*)\times\log(1-\mathcal{F}({d_i,y_c})),
\end{aligned}
\end{equation}

where $y^*=1$ if $y_c$ is the gold label of the $i^{th}$ song, $y^*=0$ if $y_c$ is a sampled negative label of the $i^{th}$ song. We replace this loss function with the nnPU loss~\cite{kiryo2017positive}, which is defined as
\begin{equation}
\begin{aligned}
    &\widetilde{R}_{pu}(\mathcal{F})=\pi_p \hat{R}_p^+ (\mathcal{F})+\max \{0, \hat{R}_u^-(\mathcal{F})-\pi_p\hat{R}_p^-(\mathcal{F})\},\\
    &\hat{R}_p^+ (\mathcal{F})=\frac{1}{|\chi_p|}\sum\limits_{d_i,y_c\in\chi_p}BCE(1, \mathcal{F}({d_i,y_c})),\\
    &\hat{R}_u^-(\mathcal{F})=\frac{1}{|\chi_u|}\sum\limits_{d_i,y_c\in\chi_u}BCE(0, \mathcal{F}({d_i,y_c})),\\
    &\hat{R}_p^-(\mathcal{F})=\frac{1}{|\chi_p|}\sum\limits_{d_i,y_c\in\chi_p}BCE(0, \mathcal{F}({d_i,y_c})),\\
\end{aligned}
\end{equation}
where $\chi_p$ and $\chi_u$ represent the positive and unlabeled samples, respectively. Besides, the nnPu loss is deployed following the implementation of \url{https://github.com/kiryor/nnPUlearning}.
\paragraph{GROOV~\cite{simig2022open}} We use the codes released by the original paper \url{https://github.com/facebookresearch/GROOV}. Particularly, for a fair comparison, we have fine-tuned the generative model on our training set.
\begin{CJK*}{UTF8}{gbsn}
\renewcommand\arraystretch{1.2}
\begin{table*}[h]
    \centering
     \caption{Chinese version of the examples in Section~\ref{sec:case}.}
     \scalebox{1.0}{
    \begin{tabular}{lp{12cm}}
    \hline\hline
    \textbf{Title: \textit{The Last Of The Mohicans}}    & \textbf{Artist: Alexandro Querevalú}\\
    \hline
     \textbf{Gold labels} & \tgold{灵魂，唯美，烦恼，激情，伤感} \\
     \hline
     \textbf{Complete set of labels} & \tfull{悠扬，骄傲，无助，人生，时代，灵魂，男性，澎湃，美洲，呐喊，未来，人间，感人，天籁，动荡，激情，伤感，民族，孤独，感动，故事，唯美，烦恼，奴隶，战争}\\
   \hline
     \textbf{LightXML~\cite{jiang2021lightxml}}    & \tgold{唯美}，\tgold{激情}，一个人，\tfull{感动}，\tfull{孤独}，\tfull{感人}，\tfull{人间}  \\
     \hline
     \textbf{EQ.~(\ref{eq:ebc})}& \tgold{激情}，\tgold{伤感}，一个人，\tfull{感动}，\tfull{故事}，\tfull{孤独}，\tfull{感人}，\tfull{人间}，\tfull{灵魂}，\tgold{唯美} \\
     \hline
     \textbf{nnPU~\cite{kiryo2017positive}  w/ EQ.~\ref{eq:ebc}}  & \tgold{激情}， \tfull{战争}，\tfull{骄傲}，\tgold{伤感}，一个人，\tfull{人生}，\tgold{故事}，哭泣，\tfull{感人}，\tfull{人间}，\tgold{未来}，视频，\tfull{时代}，\tfull{感动}，见证，和平，食，纯，希望，\tfull{天籁}，\tnew{害怕}，利益，大街，\tgold{苦难}，记住，可怜，东方 \\
     \hline
     \textbf{GROOV~\cite{simig2022open}}    & 治愈，宁静，回忆，\tnew{经典}，情歌，爱情，深情，一个人，\tfull{故事}，星河，青春，阳光，励志，热血，\tgold{唯美}，\tnew{轻音乐}，热爱，\tnew{bgm}，\tgold{灵魂}，古韵 \\
     \hline
     \textbf{ChatGPT~\cite{ouyang2022training}}   & \tnew{原生态音乐}，\tnew{民族音乐}，人文关怀，历史意义，反思人类文明，文化多样性，\tnew{悲壮情感}，\tnew{抗争与挫折} \\
     \hline
     \textbf{NST~\cite{xie2020self} w/ EQ.~(\ref{eq:ebc})}   & \tfull{人间}，\tfull{故事}，一个人，\tgold{伤感}，\tgold{激情}，\tfull{感人}，\tfull{感动}，\tfull{孤独}，\tgold{唯美} \\
     \hline
     \textbf{\name}    & \tfull{感人}，\tgold{伤感}， 一个人，\tfull{感动}，\tfull{故事}，\tfull{孤独}，\tgold{激情}，\tgold{唯美}，\tfull{动荡}，\tfull{人间}，\tfull{男性}，\tfull{奴隶}，西班牙，\tnew{吹奏}，\tfull{天籁}，\tfull{民族}，\tfull{时代}，\tnew{荡气回肠}，\tnew{震撼}，诚信，\tgold{伤感}，\tfull{战争}，民主，\tgold{灵魂}，\tfull{人生}，\tfull{澎湃}，\tfull{悠扬} \\
     \hline
     \hline
        \textbf{Title: \textit{Mermaid}}    & \textbf{Artist: Skott}\\
     \hline
     \textbf{Gold labels}    & \tgold{相见}\\
     \hline
     \textbf{Complete set of labels}    & \tfull{情深，空旷，宁静，清冷，感染力，打雷，剪辑，森林，蛊惑，空灵，相见，美好，开心，深情，海盗，故事，惊艳，爱情} \\
     \hline
     \textbf{LightXML~\cite{jiang2021lightxml}}    & \tfull{深情}，\tfull{情深}，\tfull{美好}\\
     \hline
     \textbf{EQ.~(\ref{eq:ebc})}& \tfull{爱情}，\tfull{情深}，\tfull{深情}，\tfull{故事}，\tfull{美好}\\
     \hline
     \textbf{nnPU~\cite{kiryo2017positive}  w/ EQ.~\ref{eq:ebc}}  & \tfull{深情}，\tfull{情深，故事}，视频，\tfull{美好}，\tfull{惊艳}，分享，\tfull{宁静}，\tfull{森林}，\tfull{空灵}，希望，\tfull{开心}，声线，风格，值得，期待，音色，\tgold{相见，干净}，宝藏，渴望，收藏，人心\\
     \hline
     \textbf{GROOV~\cite{simig2022open}}    & 治愈，星河，回忆，\tfull{宁静}，情歌，\tfull{深情}，\tfull{爱情}，伤感，经典，青春，阳光，\tgold{相见}，一个人，bgm，节奏，激情，灵魂\\
     \hline
     \textbf{ChatGPT~\cite{ouyang2022training}}& \tnew{知名歌手打雷姐}，\tnew{美人鱼}，海妖，\tnew{Skott}，Bjork，华为，冰雪2剪辑视频\\
     \hline
     \textbf{NST~\cite{xie2020self} w/ EQ.~(\ref{eq:ebc})}   &  \tfull{故事}，\tfull{爱情}，\tfull{情深}，\tfull{深情}，\tfull{空灵}，\tfull{森林}，\tfull{美好}， \tfull{开心}，\tfull{宁静} \\
     \hline
     \textbf{\name}    & \tfull{美好}，\tfull{爱情}，\tfull{故事}，\tfull{深情}，\tfull{情深}，\tfull{海盗，空灵}，\tfull{森林，蛊惑}，\tfull{空旷}，\tfull{清冷}，路上，\tfull{宁静，感染力}，\tfull{开心}，\tfull{惊艳}，\tgold{相见}\\
     \hline\hline

    \end{tabular}
}
    \label{tab:case-ch}
\end{table*}
\end{CJK*}
\paragraph{ChatGPT~\cite{ouyang2022training}} We have tried the prompts provided by \citet{gilardi2023chatgpt,10.1145/3543873.3587368,moller2023prompt}. To better instruct ChatGPT to perform the task, all prompts are translated into Chinese with minor changes. The results in Table~\ref{tab:main-1} are obtained from the best-performing prompt, which is
\begin{CJK*}{UTF8}{gbsn}
``请从歌曲评论$\{\}$中提取出多样的歌曲标签, 例如$\{\}$.format(user comments, gold labels)''
\end{CJK*}
(``Please select diverse labels from the user comments of song $\{\}$, for example , $\{\}$ .format(user comments, gold labels)'').
\paragraph{Noisy Student Training~(NST)~\cite{xie2020self}} The codes released by the original paper \url{https://github.com/google-research/noisystudent} are based on image classification. In this paper, we replace the image classifier with MLC and EQ.~(\ref{eq:ebc}). 
\subsection{Metrics}
Besides the metrics~(P, SR, SF1, and Coverage scores) defined Sec.~\ref{sec:setting}. We use the scikit-learn tool~(\url{https://scikit-learn.org/stable/modules/classes.html#module-sklearn.metrics}) to compute the precision, recall, F1, and nDCG scores. Moreover, we use the implementation of \url{https://github.com/kunaldahiya/pyxclib} to compute the propensity scored precision~(PSP) and propensity scored nDCG~(PSnDCG) scores. Specifically, the propensity score of a label is estimated via label priors $\pi_{y^*}:=\mathbb{P}[y^*=1]$:
\begin{equation}
    p_{y^*}:=\frac{1}{1+(\log N-1)\times (b+1)^a\times e^{-a\log (N\pi_{y^*}+b)}},
\end{equation}
where $a=0.55$ and $b=1.5$, as recommended by \citet{jain2016extreme}.

\section{CHINESE VERSION OF EXAMPLES IN Sec.~\ref{sec:case}}
Tabel~\ref{tab:case-ch} presents the original Chinese version of the examples used Sec.~\ref{sec:case}. 



\end{document}